\documentclass{aa}
\usepackage{times}
\usepackage{epsfig}
\usepackage{graphics}
\usepackage{amssymb}

\begin{document}

\title{The afterglow of the short/intermediate-duration gamma-ray burst 
       \object{GRB~000301C}:
A jet at z=2.04\thanks{Based on observations made with the 
              Nordic Optical Telescope, operated on the island of La Palma 
              jointly by Denmark, Finland, Iceland, Norway, and Sweden, in 
              the Spanish Observatorio del Roque de los Muchachos of the 
              Instituto de Astrofisica de Canarias.}\fnmsep\thanks{Based on 
              observations collected at the European Southern Observatory, 
              La Silla and Paranal, Chile (ESO project No. 64.H-0573)}
              \fnmsep\thanks{Based on observations at the German-Spanish
              Astronomical Centre, Calar Alto, operated by the
              Max-Planck-Institute for Astronomy, Heidelberg, jointly
              with the Spanish National Commission for Astronomy.}
}

\titlerunning{The Afterglow of GRB~000301C}

\author{
   B.L. Jensen \inst{1}
   \and J.U. Fynbo \inst{2,3}
   \and J. Gorosabel \inst{4} 
   \and J. Hjorth \inst{1}
   \and S. Holland \inst{2,5}
   \and P. M\o ller \inst{3}
   \and B. Thomsen \inst{2}
   \and G. Bj\"{o}rnsson \inst{6}
   \and H. Pedersen \inst{1}
   \and I. Burud \inst{7}
   \and A. Henden \inst{8}
   \and N.R. Tanvir \inst{9}
   \and C.J. Davis \inst{10}
   \and P. Vreeswijk \inst{11}
   \and E. Rol \inst{11}
   \and K. Hurley \inst{12}
   \and T. Cline \inst{13}
   \and J. Trombka \inst{14}
   \and T. McClanahan \inst{14}
   \and R. Starr \inst{15}
   \and J. Goldsten \inst{16}
   \and A.J. Castro-Tirado \inst{17,18}
   \and J. Greiner \inst{19}
   \and C.A.L. Bailer-Jones \inst{20}
   \and M. K{\"u}mmel \inst{20}
   \and R. Mundt \inst{20}
}

\institute{ 
           Astronomical Observatory, 
           University of Copenhagen, 
           Juliane Maries Vej 30, DK--2100 Copenhagen \O, Denmark;
           brian\_j@astro.ku.dk,jens@astro.ku.dk,holger@astro.ku.dk
           \and
           Institute of Physics and Astronomy,
           University of Aarhus, DK--8000 \AA rhus C., Denmark;
           jfynbo@eso.org, sholland@nd.edu,bt@ifa.au.dk
           \and
           European Southern Observatory,
           Karl--Schwarzschild--Stra\ss e 2,
           D--85748 Garching, Germany;
           pmoller@eso.org
           \and
           Danish Space Research Institute,
           Juliane Maries Vej 30, DK--2100 Copenhagen \O, Denmark;
           jgu@dsri.dk
           \and
	   Department of Physics, University
	   of Notre Dame, Notre Dame IN 46556-5670, U.S.A
	   \and
           Science Institute,
           Dunhagi 3, University of Iceland, IS--107 Reykjavik, Iceland;
           gulli@raunvis.hi.is
           \and
           Institut d'Astrophysique et de G\' eophysique, Universit\' e de
           Li\`ege, Avenue de Cointe 5, B--4000 Li\`ege, Belgium;
           burud@astro.ulg.ac.be
           \and
           U.S. Naval Observatory,
           Flagstaff Station, Flagstaff, AZ 86002--1149;
           aah@nofs.navy.mil
           \and
           Department of Physical Sciences, 
           University of Hertfordshire, College Lane, Hatfield,
           Herts AL10 9AB, UK;
           nrt@star.herts.ac.uk
           \and
           Joint Astronomy Centre, 660 N. A'ohoku Place, Hilo,
           Hawaii 96720, USA;
           c.davis@jach.hawaii.edu
           \and
           University of Amsterdam,
           Kruislaan 403, 1098 SJ Amsterdam, The Netherlands;
           pmv@astro.uva.nl,evert@astro.uva.nl
           \and
           University of California, Berkeley,
           Space Sciences Laboratory,
           Berkeley, CA 94720--7450;
           khurley@ssl.berkeley.edu
           \and
           NASA Goddard Space Flight Center,
           Code 661, Greenbelt, MD 20771;
           cline@lheavx.gsfc.nasa.gov
           \and
           NASA Goddard Space Flight Center,
           Code 691, Greenbelt, MD 20771;
           Jacob.I.Trombka.1@gsfc.nasa.gov, xrtpm@leptpm.gsfc.nasa.gov
           \and
           The Catholic University of America,
           Department of Physics,
           Washington, DC 20064;
           rstarr@lepvax.gsfc.nasa.gov
           \and
           The Johns Hopkins University,
           Applied Physics Laboratory,
           Laurel, MD 20723;
           john.goldsten@jhuapl.edu
           \and
           Laboratorio de Astrof\'{\i}sica Espacial y F\'{\i}sica Fundamental 
           (LAEFF-INTA), P.O. Box 50727, E-28080, Madrid, Spain;
           ajct@iaa.es
	   \and
	   Instituto de Astrof\'{\i}sica de Andaluc\'{\i}a (IAA-CSIC),
	   P.O. Box 03004, E-18080 Granada, Spain
           \and
           Astrophysikalisches Institut, Potsdam, Germany;
           jgreiner@aip.de
           \and
           Max-Planck-Institut f{\"u}r Astronomie,
           K{\"o}nigstuhl 17,
           D-69117 Heidelberg,
           Germany;
           calj@mpia-hd.mpg.de, Kuemmel@mpia-hd.mpg.de, mundt@mpia-hd.mpg.de
           }
\offprints{B.L. Jensen}
\mail{brian\_j@astro.ku.dk}

\date{Received  / Accepted }


\abstract{
        We present Ulysses and NEAR data from the detection of the
short or intermediate duration (2 s) gamma-ray burst
\object{GRB~000301C} (2000 March 1.41 UT).  The gamma-ray burst (GRB)
was localised by the Inter Planetary Network (IPN) and RXTE to an area
of $\sim$50 arcmin$^2$. A fading optical counterpart was subsequently
discovered with the Nordic Optical Telescope (NOT) about 42~h after
the burst. The GRB lies at the border between the long-soft and the
short-hard classes of GRBs. If \object{GRB~000301C} belongs to the
latter class, this would be the first detection of an afterglow to a
short-hard burst.  We present UBRI photometry from the time of the
discovery until 11 days after the burst. We also present JHK
photometry obtained with UKIRT on 2000 March 4.5 UT (3.1 days after
the burst). Finally, we present spectroscopic observations of the
optical afterglow obtained with the ESO VLT Antu telescope 4 and 5
days after the burst. The optical light curve is consistent with being
achromatic from 2 to 11 days after the burst and exhibits a break.  A
broken power-law fit yields a shallow pre-break decay power-law slope
of $\alpha_1=-0.72\pm$0.06, a break time of $t_{\rm
break}=4.39\pm$0.26 days after the burst, and a post-break slope of
$\alpha_2=-2.29\pm$0.17. These properties of the light curve are best
explained by a sideways expanding jet in an ambient medium of constant
mean density. In the optical spectrum we find absorption features that are
consistent with \ion{Fe}{ii}, \ion{C}{iv}, \ion{C}{ii}, \ion{Si}{ii}
and Ly$\alpha$ at a redshift of 2.0404$\pm$0.0008. We find evidence
for a curved shape of the spectral energy distribution of the observed
afterglow.  It is best fitted with a power-law spectral distribution
with index $\beta\sim -0.7$ reddened by an \object{SMC}-like
extinction law
with A$_{V}\sim 0.1$ mag.  Based on the Ly$\alpha$ absorption line
we estimate the \ion{H}{i} column density to be
$\log$(N(\ion{H}{i}))$=21.2\pm0.5$.  This is the first direct
indication of a connection between GRB host galaxies and Damped
Ly$\alpha$ Absorbers.
}

\maketitle


\section{Introduction}

        The discovery of the first $X$-ray afterglow (Costa
et~al.~\cite{C1997}) and optical counterpart (van~Paradijs
et~al.~\cite{PGG1997}) to a long-duration gamma-ray burst (GRB) have
led to a revolution in GRB research. The determination of a redshift
of 0.835 for \object{GRB~970508} (Metzger et~al.~\cite{MDK1997}), and
the subsequent determination of redshifts of 13 bursts with a median
redshift of $\sim$1.0, have firmly established their cosmological
origin (Kulkarni et al.~\cite{KBB2000}; This work; Bloom et al.~\cite{BDD2000}).

        The intriguing case of an association of the peculiar
supernova \object{SN1998bw} with \object{GRB~980425} (Galama
et~al.~\cite{GVP1998}) was the first indication of a possible
connection with supernovae.
Evidence for supernova signatures in the late light curves of 
\object{GRB~970228} (Reichart~\cite{Rei1999}; Galama~et~al.~\cite{GTV1999})
and \object{GRB~980326} (Castro-Tirado \& Gorosabel~\cite{CTG1999}; Bloom~et~al.~\cite{BKD1999}) suggests that at least some long-duration GRBs may be related to the collapse of massive ($>25$ M$_\odot$) stars.
Breaks in the power-law declines of \object{GRB~990123}
(Kulkarni~et~al.~\cite{KDO1999}) and \object{GRB~990510} 
(Harrison~et~al.~\cite{HBF1999}) are interpreted as evidence for collimated 
outflows (`jets') (see also Holland et al.~\cite{HBH2000}). Further evidence 
for this collapsar + jet model 
(e.g., MacFadyen \& Woosley~\cite{MW1999}) comes from the light curve of
\object{GRB~980519} which is best interpreted as a jet expanding into a 
preexisting circumburst stellar wind (Jaunsen et~al.~\cite{J2001}).

        The high-energy properties of GRBs show a bi-modal distribution of
burst durations (Kouveliotou et~al.~\cite{K1995}) which, in the
simplest scenario, may indicate the existence of binary compact mergers
as the progenitors of the short-duration bursts (T$_{90}<2$ s).
From an analysis of the Third BATSE Catalog, Mukherjee et
al.~(\cite{MFB1998}) have shown that, in addition to the short
(T$_{90}<2$ s) and long (T$_{90}>5$ s) classes, there may exist a
third, intermediate soft-spectrum class of GRBs with duration 2 $\mathrm{s} <
\mathrm{T}_{90} < 5$ s.

        In this paper we report the discovery and subsequent observations and
analysis of the afterglow of the short-to-intermediate duration
\object{GRB~000301C} (Fynbo et~al.~\cite{FJH2000a}).

        Sect.~\ref{SECTION:detection} reports the detection, IPN localisation
and the high-energy data of the GRB obtained from Ulysses and
NEAR\@. Sect.~\ref{SECTION:discovery} describes the discovery of the
optical counterpart and our subsequent optical and infrared
observations.  Sect.~\ref{SECTION:photometry} details the optical and
infrared photometry and Sect.~\ref{SECTION:Spectroscopy} describes the
VLT spectroscopy.  Sect.~\ref{SECTION:Results} describes the results
obtained on the spectroscopy and spectral energy distribution
and Sect.~\ref{SECTION:discussion} is devoted to the
discussion and interpretation, with Sect.~\ref{SECTION:conclusion}
presenting our conclusions.

        Throughout this paper, we adopt a Hubble constant of H$_0$ = 65 km
s$^{-1}$ Mpc$^{-1}$ and assume $\Omega_{\mathrm{m}}=0.3$ and $\Omega_{\Lambda}=0.7$.


\section{Detection and localisation of the gamma-ray burst}
\label{SECTION:detection}

        \object{GRB~000301C} was recorded by the Ulysses GRB experiment and by
the NEAR $X$-Ray/Gamma-Ray Spectrometer. Because this burst was relatively weak, it did not trigger the Ulysses Burst Mode, and the only data available from Ulysses is the Observation Mode 1 0.25~s resolution 25--150~keV light curve (Hurley et al.~\cite{HSA1992}).  NEAR records the light curves of bursts in the 150--1000~keV energy range with 1~second resolution, but takes high-energy spectra only with 40~min resolution.

        Analysis of the Ulysses and NEAR relative timing data yields an
annulus centred at
$(\alpha,\delta)_{2000}=(20^{\mathrm{h}}34^{\mathrm{m}}7.56^{\mathrm{s}},
+20^{\circ}32{\arcmin}19.62{\arcsec})$, with a radius of
57.520$\pm0.083$ degrees (at 3$\sigma$ full-width).  This annulus
intersected the error-box of the All-Sky Monitor (ASM) on the RXTE
spacecraft, at near-right angles to create a composite localisation of
a parallelogram of area $50$ arcmin$^2$ (see Fig.~\ref{FIGURE:FC}).

Since no high-energy spectra are available, we have estimated the peak fluxes 
and fluences for trial power-law spectra with indices between 1 and 4 using 
the Ulysses data.  For a typical power-law index of 2, we find a 25--100~keV 
fluence of $2.1 \times 10^{-6} \,\mathrm{erg\, cm}^{-2}$, and a peak flux over 
the same energy range, and over 0.25~s, of $6.3 \times 10^{-7}\, 
\mathrm{erg\, cm}^{-2}\, s^{-1}$.
The uncertainties in these numbers are partly due to 
the lack of a high-energy spectrum.
For example, the fluence estimates range from 
$1.45 \times 10^{-6}$ to $2.24 \times 10^{-6}\, \mathrm{erg\, cm}^{-2}$ as the 
spectral index is varied from 4 to 1. The statistical uncertainty is 
approximately 30\%. From the NEAR data we estimate the 150--1000~keV fluence 
to be approximately $2 \times 10^{-6}\, \mathrm{erg\, cm}^{-2}$.

To date, the only GRBs with identified long-wavelength counterparts have been 
long-duration bursts.  As measured by both Ulysses and NEAR, in the $>$25~keV
energy range, the duration of this burst was approximately 2 s. 
(Note that the earlier estimate of a 10 s duration of \object{GRB~000301C} by 
Smith et al.~(\cite{SHC2000}) was based on the $<$10~keV energy range).  Thus 
it falls in the short class of bursts, though it is consistent with belonging 
to the proposed intermediate class or the extreme short end of the distribution
of long-duration GRBs (Hurley et al.~\cite{H1992}; Mukherjee et 
al.~\cite{MFB1998}). Although we do not have any measurements of 
the high-energy spectra above 25~keV, it is possible to derive a crude estimate
of the spectral index, and therefore the hardness ratio (the 100--300 keV 
fluence divided by the 50--100 keV fluence), from the Ulysses and NEAR count 
rates.  We obtain a hardness ratio of 2.7$\pm$0.6(cutoff)$\pm$30\%(statistical 
error) from fitting a power-law, with the index as a free parameter, to the
count rates from NEAR and Ulysses, assuming a range of cut-off energies.

\begin{figure}
\epsfig{file=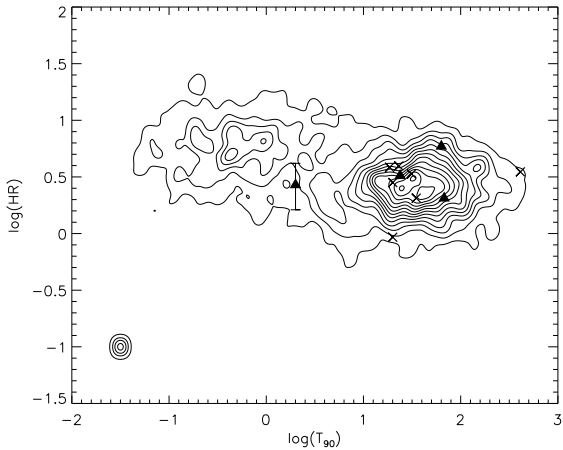,width=8.0cm}
\caption{A contour plot showing the duration--hardness
($\log$(T$_{90}$)--$\log$(HR$_{32}$)) distribution of BATSE bursts
(from the Fourth BATSE GRB Catalog (revised) (Paciesas et
al.~\protect\cite{PMP1999}) and the Current BATSE GRB Catalog). The triangle
with an error-bar near the center of the plot represents
\object{GRB~000301C}. Other symbols represent 10 other BATSE bursts
with identified counterparts for which data on fluence and duration
are available. Triangles are bursts which have a break in their
optical light curves. Errors in the BATSE data are smaller than the
symbol size. Contour levels scale linearly. The centroid in the lower left
corner indicates the resolution.}

\label{FIGURE:grb_classes}
\end{figure}

        Fig.~\ref{FIGURE:grb_classes} shows the location of
\object{GRB~000301C} in a hardness vs.\ duration plot.  The contour
plot contains 1959 GRBs for which data on fluence and duration were
available in the Fourth BATSE GRB Catalog (revised) (Paciesas et
al.~\cite{PMP1999}) and the BATSE Current GRB Catalog\footnote{Data on
current GRBs are available through the BATSE homepage {\tt \small
http://www.batse.msfc.nasa.gov/batse/}}. The symbols represent the 10
GRBs included in the BATSE catalogs for which an afterglow has been
identified, with \object{GRB~000301C} located near the center of the
plot. Triangles are bursts where a break has been found in the
optical light curve. From this sparse set of data there does not
appear to be any marked difference in the distributions of bursts
with, or without, an identified break.

Of the 1959 BATSE bursts in Fig.~\ref{FIGURE:grb_classes}, the ratio between 
bursts with a duration of T$_{90} \geq 2.0$ s and with T$_{90} < 2.0$ s is 3:1.
To date, at least 23 GRB optical afterglows have been
discovered (Kulkarni et al.~\cite{KBB2000}; Andersen et
al.~\cite{AH2000}; this work; Klose et~al.~\cite{KSF2000}; Fynbo et~al.~\cite{FJG2001}; Fynbo et al.~\cite{FGD2001}; Henden \cite{H2001}; grb-webpage of J. Greiner\footnote{\tt http://www.aip.de/\~{ }jcg/grbgen.html}).
 If the distribution of the 23 GRBs with identified counterparts follows the
general BATSE distribution, one would expect that $17\pm4$ bursts were
in the long class, and $6\pm2$ bursts were in the short class.
However, \object{GRB~000301C} is the only GRB with a duration
consistent with the short-duration class. The expected number of
identified short burst counterparts is moderated by the strong
selection bias caused by the technical difficulties of obtaining
precise localisations for the short GRBs.

\section{Discovery and observations of the afterglow}
\label{SECTION:discovery}
\begin{figure}
\epsfig{file=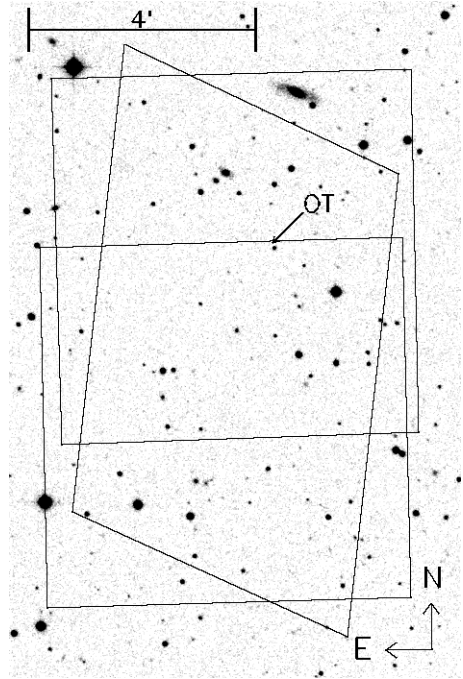,width=8.0cm}
\caption{A finding chart with the rectangular IPN-errorbox indicated
along with the two initial ALFOSC pointings (ALFOSC FOV = 6.5~\arcmin).}
\label{FIGURE:FC}
\end{figure}

The IPN/RXTE error-box of \object{GRB~000301C} (Smith~et~al.~\cite{SHC2000})
was observed with the 2.56-m Nordic Optical Telescope (NOT) on 2000 
March 3.14--3.28 UT ($\sim$1.8 days after the burst) using the 
Andaluc\'{\i}a Faint Object Spectrograph (ALFOSC). Comparing with red and blue 
Palomar Optical Sky Survey~II exposures, a candidate Optical Transient (OT) 
was found at the position
$(\alpha,\delta)_{2000}=(16^{\mathrm{h}}20^{\mathrm{m}}18.56^{\mathrm{s}}, 
+29^{\circ}26{\arcmin}36.1{\arcsec})$ (Fynbo et al.~\cite{FJH2000a}).
A finding-chart of the IPN-errorbox and the two ALFOSC pointings 
used to cover the field can be seen in Fig.~\ref{FIGURE:FC}.

\begin{figure}
\epsfig{file=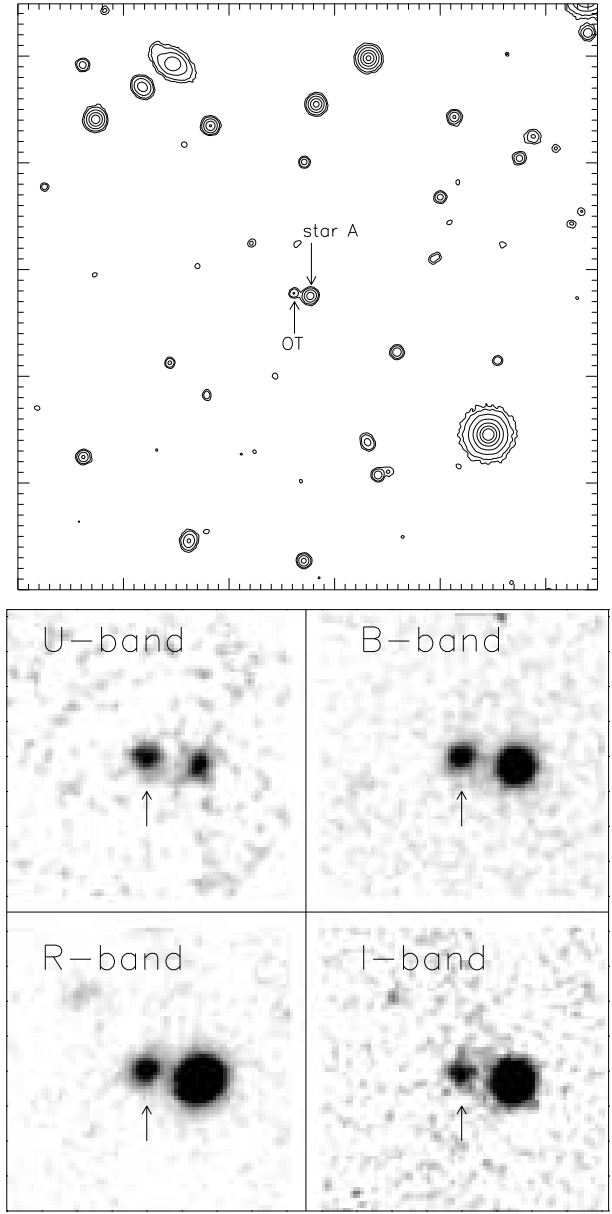,width=8.0cm}
\caption{{\it Upper panel\/}: A contour plot showing a 3.5$\times$3.5
arcmin$^2$ region centred on the Optical Transient (OT) of
\object{GRB~000301C} from the combined R-band image obtained at the
NOT on 2000, March 3 UT\@. The OT is located 6 arcsec east of the star
marked A (following Garnavich et al.~\protect\cite{GBJ2000}). {\it Lower panel\/}: A
region of 30$\times$30 arcsec$^2$ centred on the OT from the combined
U, B, R and I frames obtained on 2000, March 3.14--3.28 UT, about 42
hours after the burst.}
\label{FIGURE:OT}
\end{figure}

The transient nature of the candidate was subsequently confirmed 
at optical and infrared wavelengths (Bernabei et al.~\cite{BMB2000}; 
Stecklum et al.~\cite{SKF2000}; Garnavich et al.~\cite{GBJ2000}; 
Veillet et al.~\cite{V2000}; Fynbo et al.~\cite{FJH2000b}; 
Kobayashi et al.~\cite{KGT2000}). At the time of discovery the magnitude of 
the OT was R=20.09$\pm$0.04 (see Sect.~\ref{SECTION:photometry} for a detailed 
discussion of the photometry and Fig.~\ref{FIGURE:OT} for a finding chart and 
UBRI-images of the OT on 2000 March 3 UT). 

We obtained subsequent optical observations using the NOT, the Antu
telescope (UT1) of ESO's Very Large Telescope (VLT), the USNOFS 1.0-m
telescope, the 2.2-m telescope at Calar Alto (CAHA) and the Wide Field
Imager (WFI) on the ESO 2.2-m telescope. In addition we obtained
near-infrared (NIR) data from the United Kingdom Infra-Red Telescope
(UKIRT).

        The journal of our optical and NIR observations, including the
derived magnitudes, is given in Table~\ref{TABLE:photometry}.

The OT was also observed from several other optical and infrared telescopes, 
and the counterpart was subsequently detected in the mm-band at 250~GHz (Bertoldi \cite{B2000}), and at radio (8.46~GHz) wavelengths 
(Berger \& Frail~\cite{BF2000})\footnote{See the GCN Circulars Archive for 
further details ({\tt \small http://gcn.gsfc.nasa.gov/gcn/gcn3\_archive.html})}.
Papers detailing the properties of the counterpart of 
\object{GRB~000301C} have been presented at radio and mm wavelengths 
(Berger et al.~\cite{BSF2000}) and
in the infrared (Rhoads \& Fruchter~\cite{RF2001}), 
and optical bands (Masetti et al.~\cite{MBB2000}; Sagar et al.~\cite{SMP2000}).

\begin{table}
\begin{center}
\caption{Journal of our observations of \object{GRB~000301C}
with NOT+ALFOSC, USNO, CAHA 2.2-m, ESO VLT+FORS1, ESO 2.2-m+WFI in the optical bands, and the UKIRT observations in the infrared bands. The magnitudes were obtained from PSF photometry of the OT using DAOPHOT~II, except for the VLT observation of 2000 March 6.39 UT, which has been derived from the March 6 VLT spectra (see Table~\ref{TABLE:spectroscopy}). }
{\scriptsize
\begin{tabular}{@{}llcccr}
Telescope & date       & seeing & filter & exp.\ time & mag  \\
          & (2000, UT) & arcsec &        &  (sec)    &      \\
\hline
NOT       &  Mar 3.14  &  2.3   &   R    &  900   & 20.09$\pm$0.04 \\
NOT       &  Mar 3.17  &  2.2   &   R    &  900   & 20.15$\pm$0.04 \\
NOT       &  Mar 3.19  &  2.1   &   R    &  900   & 20.11$\pm$0.04 \\
NOT       &  Mar 3.20  &  2.2   &   B    &  900   & 21.05$\pm$0.04 \\
NOT       &  Mar 3.21  &  2.3   &   R    &  900   & 20.14$\pm$0.04 \\
NOT       &  Mar 3.22  &  2.3   &   B    &  900   & 21.02$\pm$0.04 \\
NOT       &  Mar 3.24  &  2.3   &   U    &  900   & 20.52$\pm$0.08 \\
NOT       &  Mar 3.25  &  2.2   &   R    &  900   & 20.16$\pm$0.04 \\
NOT       &  Mar 3.26  &  2.0   &   I    &  900   & 19.60$\pm$0.07 \\
NOT       &  Mar 3.27  &  2.3   &   I    &  300   & 19.59$\pm$0.06 \\
\hline
CAHA 2.2m &  Mar 4.14  &  1.4   &   I    &  1440   & 20.10$\pm$0.10 \\
\hline
USNO      &  Mar 4.39  &  2.0   &   I    &  480    & 20.04$\pm$0.11 \\
USNO      &  Mar 4.40  &  2.1   &   V    &  720    & 21.05$\pm$0.07 \\
USNO      &  Mar 4.41  &  2.0   &   B    &  1200   & 21.35$\pm$0.05 \\
USNO      &  Mar 4.42  &  1.8   &   R    &  600    & 20.61$\pm$0.06 \\
USNO      &  Mar 4.48  &  1.6   &   R    &  1200   & 20.58$\pm$0.03 \\
USNO      &  Mar 4.49  &  1.6   &   R    &  1200   & 20.54$\pm$0.04 \\
USNO      &  Mar 4.50  &  1.6   &   R    &  1200   & 20.60$\pm$0.04 \\
USNO      &  Mar 4.52  &  1.6   &   B    &  1200   & 21.42$\pm$0.04 \\
USNO      &  Mar 4.53  &  1.6   &   V    &  1200   & 20.98$\pm$0.04 \\
USNO      &  Mar 4.43  &  2.0   &   U    &  1200   & 20.82$\pm$0.11 \\
\hline
UKIRT     &  Mar 4.51  &  0.7   &   J    &  810    & 19.19$\pm$0.04 \\
UKIRT     &  Mar 4.53  &  0.6   &   H    &  810    & 18.44$\pm$0.04 \\
UKIRT     &  Mar 4.55  &  0.5   &   K    &  810    & 17.62$\pm$0.05 \\
\hline
CAHA 2.2m &  Mar 5.23  &  1.5   &   I    &  960    & 20.24$\pm$0.12 \\
\hline
ESO VLT   &  Mar 5.39  &  0.9   &   R    &  10     & 20.61$\pm$0.05 \\
\hline
ESO VLT   &  Mar 6.39  &  1.1   &   R    &  ...    & 21.43$\pm$0.26 \\
\hline
NOT       &  Mar 7.22  &  1.4   &   R    &  300    & 21.59$\pm$0.07 \\
NOT       &  Mar 7.24  &  1.4   &   U    &  3600   & 21.92$\pm$0.07 \\
\hline
CAHA 2.2m &  Mar 7.23  &  1.1   &   I    &  960    & 21.13$\pm$0.15 \\
\hline
NOT       &  Mar 8.18  &  1.4   &   R    &  1200   & 21.80$\pm$0.05 \\
NOT       &  Mar 8.20  &  1.5   &   I    &  900    & 21.39$\pm$0.09 \\
NOT       &  Mar 8.21  &  1.6   &   B    &  1100   & 22.89$\pm$0.08 \\
NOT       &  Mar 8.25  &  1.9   &   U    &  3700   & 22.49$\pm$0.12 \\  
\hline
NOT       &  Mar 9.15  &  1.4   &   R    &  2000   & 22.11$\pm$0.15 \\
NOT       &  Mar 9.20  &  1.6   &   U    &  5400   & 22.69$\pm$0.20 \\
NOT       &  Mar 9.24  &  1.4   &   B    &  1200   & 23.01$\pm$0.11 \\
NOT       &  Mar 9.26  &  1.2   &   I    &  900    & 21.72$\pm$0.15 \\
\hline
CAHA 2.2m &  Mar 10.05 &  1.3   &   I   &  960    & 21.85$\pm$0.20 \\
\hline
USNO      &  Mar 10.40 &  3.0   &   B   &  1800   & 23.40$\pm$0.30 \\
\hline
CAHA 2.2m &  Mar 11.21 &  1.2   &   I   &  1440   & 22.63$\pm$0.27 \\
\hline
ESO 2.2m  &  Mar 11.39 &  2.4   &   R   &  3200   & 23.12$\pm$0.18 \\
\hline    
USNO      &  Mar 12.44 &  1.9   &   R   &  4800   & 23.10$\pm$0.22 \\
\hline
\label{TABLE:photometry}
\end{tabular}
}
\end{center}
\end{table}

\section{Photometry}
\label{SECTION:photometry}
\subsection{Optical data}
\label{SECTION:optical_data}
To avoid contamination from the nearby star A (located at a separation
of $6\arcsec$ west and $1\arcsec$ south of the OT in
Fig.~\ref{FIGURE:OT}), we measured the magnitude of the OT relative to
stars in the field by performing Point Spread Function (PSF)
photometry, using DAOPHOT~II (Stetson~\cite{S1987},~\cite{S1997}).
There are several bright and
unsaturated stars in the field from which a good PSF could be
determined.

For the data presented here, there is no indication of a contribution from a host galaxy to the emission at the position of the OT (see Sect.~\ref{SECTION:host} for a discussion of the host galaxy). Hence, extended emission from a faint galaxy at the position of the OT will not affect the PSF photometry appreciably (much less than observational errors).
The quality of the PSF photo\-metry was checked by subtracting the PSFs from the images of star A and the OT\@.  In all frames the residuals are consistent with being shot-noise. 

To avoid errors due to colour terms or colour differences
in our photometry (the conditions during most
of the observations at the NOT were possibly non-photometric due to increasing 
amounts of Saharan dust in the atmosphere over the telescope at the time of 
observations), the magnitudes of the OT for all our optical photometry were 
calibrated relative to stars of similar colours in the field.

The photometric standard UBVR$_\mathrm{C}$I$_\mathrm{C}$ calibration
of the field was performed at the USNOFS 1.0-m telescope and is
available in Henden~(\cite{H2000}).  This calibration has an
estimated zero-point uncertainty of 2 percent, which is well below the
errors in the relative magnitudes.  The results of the PSF photometry
are presented in Table~\ref{TABLE:photometry}.  The 2000 March 6.39
VLT R-data point has been derived from the March 6 combined
VLT-spectrum (Table~\ref{TABLE:spectroscopy}).

Based on this photometric calibration we conclude that star A showed no sign 
of variability within observational errors throughout our observations, and
that it had the following magnitudes:
U~=~20.427$\pm$0.133, B~=~19.837$\pm$0.030, 
V~=~18.767$\pm$0.018, R~=~18.084$\pm$0.043, and 
I~=~17.526$\pm$0.044.

\subsection{Near-infrared data}
\label{SECTION:IRdata}

        The UKIRT images were processed using the ORAC imaging data
reduction routines developed for UKIRT (Bridger et al.~\cite{BWE2000}).
The J, H and K magnitudes of the OT were then measured from the UKIRT
data as follows. First we measured the magnitude of the OT relative to
star A using DAOPHOT~II PSF photometry as described above. Then, in order to
transform this magnitude to the standard UKIRT system, we performed
aperture photometry in an aperture with a diameter of $2\farcs7$ on
calibration images obtained of the standard stars
\object{S868-G} and \object{p389-d}
from the list of UKIRT faint standards\footnote{\tt \small
http://www.jach.hawaii.edu/JACpublic/UKIRT/
astronomy/calib/faint\_stds.html} and on star A. The estimated error
in the zero-point is about 0.05 in each of J, H and K.  We have
assumed negligible extinction difference between standard and program
field. The results of these measurements are presented in
Table~\ref{TABLE:photometry}.

\section{Spectroscopy}
\label{SECTION:Spectroscopy}

Spectroscopic observations were carried out on 2000 March 5 and 6 UT with
VLT-Antu equipped with FORS1 (for details see the observing log in 
Table~\ref{TABLE:spectroscopy}).
  We used the GRIS\_300V+10 grism and the 
GG375 order separation filter, which provide a spectral coverage from 
3600~{\AA} to 8220~{\AA}  and a dispersion of 2.64~{\AA}/pixel.
The effective exposure time was 800~s on March 5.39 UT and 1200~s on 
March 6.38 UT\@. Standard procedures were used for bias and flat field
correction, and the optimal extraction procedure for faint spectra
(described in M{\o}ller~(\cite{M2000})) was used to extract one dimensional
spectra.

\begin{table}
\begin{center}
\caption{Spectroscopy of the afterglow of \object{GRB~000301C}, 
obtained with VLT+FORS1 on 2000 March 5 and 6 UT.}
\label{TABLE:spectroscopy}
\begin{tabular}{ccccc}
\hline
      Date         &  Exposure time & Airmass &  Slit width &\\
 (2000 UT)   &        (s)     &         &  ($\arcsec$)&\\
\hline
Mar 5.39 & 400 & 1.732--1.745 & 1.3 \\
Mar 5.40 & 400 & 1.719--1.729 & 1.3 \\
\hline
Mar 6.38 & 600  & 1.772--1.801 & 1.0 \\
Mar 6.39 & 600  & 1.745--1.767 & 1.0 \\
\hline
\end{tabular}
\end{center}
\end{table}

The position angle of the long slit was chosen such that both the OT and 
star A were centred onto the slit.  From the magnitude of star A we calibrated the flux of the trace of the optical counterpart.
The spectral flux calibration derived for the OT on March 5 is consistent with the optical photometry displayed in Table~\ref{TABLE:photometry}.
We derive a value for the spectral index of $\beta=-1.15\pm0.26$ on 
March 5 and $\beta=-1.43\pm0.28$ on March 6, corrected for interstellar 
extinction of $\mathrm{E}(\mathrm{B}-\mathrm{V})=0.053\pm0.020$, using the 
dust maps of Schlegel~et~al.~(\cite{SFD1998}).

\begin{figure}
\epsfig{file=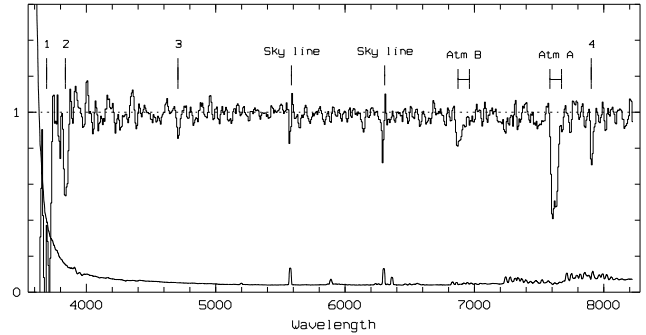, height=9.1cm, angle=270, clip=}
\caption{Combined VLT+FORS1 spectrum of \object{GRB~000301C} from 2000
March 5+6 UT\@. The spectrum is normalized to 1 in the continuum. The
atmospheric absorption bands and residuals from strong sky--lines are
marked, as well as the 4 absorption lines listed in
Table~\ref{TABLE:abs_lines}. The spectrum is binned to 7 {\AA} pixels,
and the lower curve shows the noise (per pixel).}
\label{FIGURE:Spectrum}
\end{figure}

\section{Results}
\label{SECTION:Results}

\subsection{GRB absorption lines and redshift determination}
\label{SECTION:redshift}

The combined spectrum, reproduced in Fig 3, has a resolution of
14~{\AA} FWHM, and signal--to--noise (S/N) in the range 15--30 per
resolution element redwards of 4000~{\AA}. From 4000~{\AA} to
3600~{\AA} the S/N drops rapidly.

Due to the poor resolution, only very strong absorption lines can be
detected individually. In Table~\ref{TABLE:abs_lines} we list the only
four absorption features which were detected at a S/N in excess of
4.5. Two of the features were found bluewards of 4000~{\AA}, and were
initially ignored. The line at 4712~{\AA} is broader than the
resolution profile, and we tentatively identified it as a possible
\ion{C}{iv} absorption complex with redshifts in the range 2.038 to
2.042. With this identification, the other three features would fit
the proposed identifications given in
Table~\ref{TABLE:abs_lines}. Note, however, that the \ion{Si}{ii}~1260
line is far too strong and wide to be a single line, and we hence
assume that it is a blended feature. The \ion{Fe}{ii}~2600 line is
strong and narrow, but also this line seems excessively strong given
the lack of other strong \ion{Fe}{ii} lines.

\begin{table}[t]
\begin{center}
\caption{Absorption features detected at S/N$>$4.5 in the spectrum of
\object{GRB~000301C} on 2000 March 5+6 UT\@. W$_{\rm obs}$ indicates
the observer frame equivalent width of the features.}
\label{TABLE:abs_lines}
\begin{tabular}{lccrr}
\hline
Line & Identification  & $\lambda_{\rm vac}$ & W$_{\rm obs}$ & $\sigma$(W)\\
                       &  ({\AA})            &  ({\AA})  &   &            \\
\hline
1 & Ly$\alpha$  & 3693.56 & 67.07 & 12.09 \\
2 & \ion{Si}{ii}~1260 & 3843.09 & 18.99 & 2.03  \\
3 & \ion{C}{iv}~1549  & 4712.23 &  3.76 & 0.82  \\
4 & \ion{Fe}{ii}~2600 & 7909.03 &  6.29 & 1.21  \\
\hline
\end{tabular}
\end{center}
\end{table}

In order to provide a more strict test of our proposed identification,
and to obtain an accurate value for the redshift, we proceeded as
follows. First we shifted and stacked pieces of the spectra where we
would expect common low ionization lines. We selected the singly
ionized species of Si, C and Fe, all of which are known as strong
absorbers in quasar absorption line systems. In total our spectrum
covers positions of the lines
\ion{Fe}{ii}~1608, \ion{Fe}{ii}~2344, \ion{Fe}{ii}~2374, \ion{Fe}{ii}~2382,
\ion{Fe}{ii}~2586, \ion{Fe}{ii}~2600, \ion{Si}{ii}~1260, \ion{Si}{ii}~1304,
\ion{Si}{ii}~1526, and \ion{C}{ii}~1334.
\ion{Si}{ii}~1260 (at 3832~{\AA}) was in the very low S/N part of
spectrum, and almost certainly blended, so it was not included.
Treating each ion separately, a weighted mean absorption feature was
calculated using the oscillator strength of each line as statistical
weight.

\begin{figure}
\epsfig{file=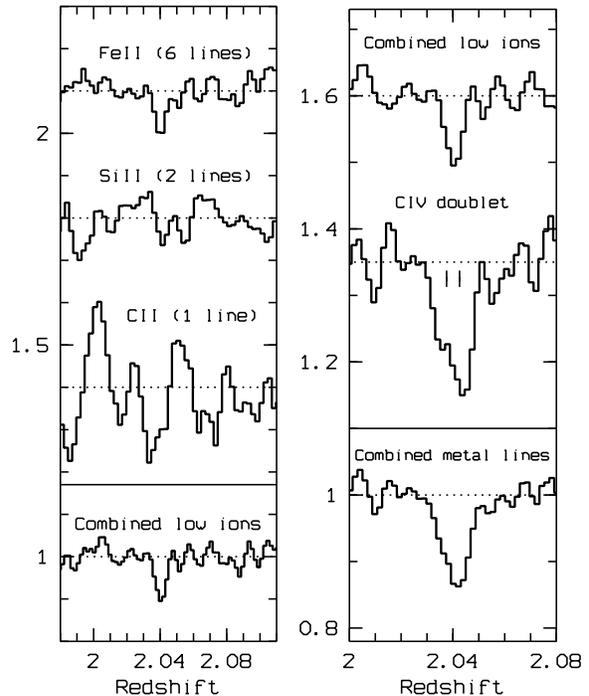, height=17.1cm, clip=}
\vskip -5.5 cm
\caption{Left: {\it Upper panel\/}: Oscillator strength weighted mean
absorption lines of \ion{Fe}{ii}, \ion{Si}{ii} and \ion{C}{ii}, lined
up in redshift space for easy comparison (see text for details).  {\it
Lower panel\/}: The combined ``Low Ionization'' absorption feature
(weighted mean of the three sections of spectrum shown above).  Right:
{\it Upper panel\/}: The combined ``Low Ionization'' absorption
feature lined up with the \ion{C}{iv} trough. {\it Lower panel\/}:
Combined ``Metal Absorption Feature''.  All the sections of spectra
have been normalized to 1 in the continuum, but shifted along the
abscissa by suitable off--sets.  Note that none of the figures start
at 0, instead the scale on the abscissa provides the proper reference.
}
\label{FIGURE:abs_troughs}
\end{figure}

The regions of the co--added spectra for each of the ions \ion{Fe}{ii},
\ion{Si}{ii} and \ion{C}{ii}, transformed into redshift space,
is shown in the left panel of Fig.~\ref{FIGURE:abs_troughs}.
A combined ``Low Ionization'' absorption feature (bottom of left
panel of Fig.~\ref{FIGURE:abs_troughs}) was obtained by co--addition
of the three sets of features, but using the number of lines
as statistical weights.
The redshift range searched for low ionization absorption systems by
this method was 1.95--2.14 and no other candidate systems were found
in this range. For comparison of the redshifts, we plot in
the right panel of Fig.~\ref{FIGURE:abs_troughs} again the
combined ``Low Ionization'' absorption feature together with
the \ion{C}{iv} absorption trough. The bottom panel here
shows the combination of all the lines. Given the significance
of this combined ``Metal Absorption Feature'', we conclude that
the tentative identification of this system is confirmed.

It is commonly seen in quasar absorption line systems that the low
ionization species, tracing the cold dense gas, have a more
well-defined redshift than the high ionization species. Hence, we shall
adopt the redshift $z_{\rm abs} = 2.0404 \pm 0.0008$, measured from
the combined Low Ionization feature, as the systemic redshift. It is seen
from the bottom right panel of Fig.~\ref{FIGURE:abs_troughs} that
inclusion of \ion{C}{iv} would result in a slightly higher
redshift. Note that $z_{\rm abs} = 2.0404 \pm 0.0008$ is consistent
with the redshift z=1.95$\pm0.1$ based on the Lyman break 
(Smette et al.~\cite{SFG2000}; Feng et al.~\cite{FWW2000}),
but is significantly higher than
the value z=2.0335$\pm$0.0003 reported by Castro et al.~(\cite{CDD2000}).

The oscillator strength weighted mean observed equivalent width of the
\ion{Fe}{ii} lines is 2.56~{\AA}, which is strong enough that by
comparison to known quasar absorbers one would expect this to likely
have a column density of neutral Hydrogen in excess of $2\times
10^{20} \mathrm{cm}^{-2}$. Such absorbers are known as Damped
Ly$\alpha$ Absorbers (DLAs), and hold a special interest because of
the large amounts of cold gas locked up in those objects (Wolfe et
al.~\cite{WLF1995}; Storrie-Lombardi et al.~\cite{SLI1997}).  It is
commonly assumed that the DLAs are the progenitors of present day disk
galaxies, but they have proven extremely difficult to identify (see
e.g.  M{\o}ller \& Warren 1993; Kulkarni et al.~\cite{KHS2000}).
Observational evidence has been accumulating (M{\o}ller \& Warren
1998; Fynbo et al.~\cite{FMW1999}) which suggests that a likely reason
why DLA galaxies are so hard to identify is their small gas
cross--sections and faint magnitudes, causing them to stay hidden
under the point spread functions of the bright quasars. A GRB selected
DLA galaxy sample would not be hampered by this problem once the OT
has faded sufficiently, and could as such help greatly in
understanding the nature of DLA galaxies. We shall therefore now
briefly consider the low S/N part of the GRB spectrum below
4000~{\AA}, to investigate if any information concerning the
\ion{H}{i} column density can be extracted.

\begin{figure}
\epsfig{file=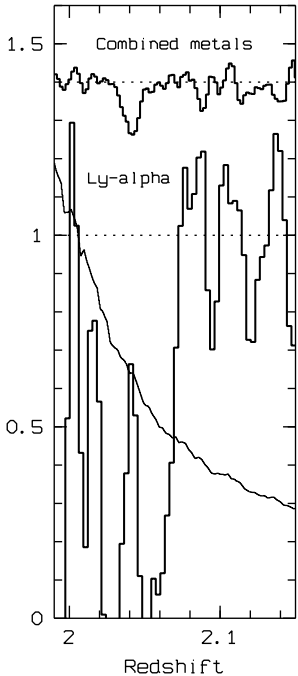, height=15.1cm, angle=270, clip=}
\vskip -1.0 cm
\caption{{\it Upper panel\/}: The ``Combined Metals'' absorption
feature from Fig.~\ref{FIGURE:abs_troughs} is shown here for comparison.
{\it Lower panel\/}: Spectral region
around Ly$\alpha$ (for comparison to the metal absorption also
plotted in redshift space).  Note the very sharp onset of absorption
well above the expected redshift. This is consistent with a very broad
Ly$\alpha$ absorption line as detailed in the text.
The lower curve shows the noise per bin (bins of 0.003 in redshift).}
\label{FIGURE:lyalpha}
\end{figure}

In Fig.~\ref{FIGURE:lyalpha} (lower panel) we have plotted the
spectral region around Ly$\alpha$, and for comparison of redshifts,
the ``Combined Metals'' feature (upper panel). Also plotted on the
lower panel is the noise per bin (for redshift binsize = 0.003).
It is clearly seen that the spectrum drops steeply before the
expected central position of the Ly$\alpha$ line,
and well before the S/N drops below detection. One likely explanation
for this is the pre\-sence of a very broad Ly$\alpha$ absorption
line. To quantify this we have modelled several Ly$\alpha$ absorption
lines, all at redshift 2.0404, and calculated the $\chi^2$ of their
fit to the data in the range 3700~{\AA} to 3750~{\AA}.
For N$(\ion{H}{i}) = 0$ the $\chi^2$ per degree-of-freedom DOF is 6.46 which
confirms that an absorption feature is indeed present.
The formal $\chi^2$ minimum is found at N$(\ion{H}{i}) =
1.5\times 10^{21} \mathrm{cm}^{-2}$ ($\chi^2$ per DOF = 0.86),
but any value within a factor 3 of this is acceptable.

It should be recalled that the above estimate, in a strict
sense, only applies in the case the OT lies ``behind'' the absorbing
cloud. In case the OT is in fact embedded ``inside'' a DLA cloud,
resonant scattering of Ly$\alpha$ photons may alter the
profile of the absorption line somewhat. In the present case this
is a detail which the quality of our data will not allow us to
discern.

\subsection{The multi-wavelength spectrum around March 4.5 UT}
\label{SECTION:SpectroscopyRes}

\begin{figure*}
\resizebox{\hsize}{!}{\includegraphics[angle=270]{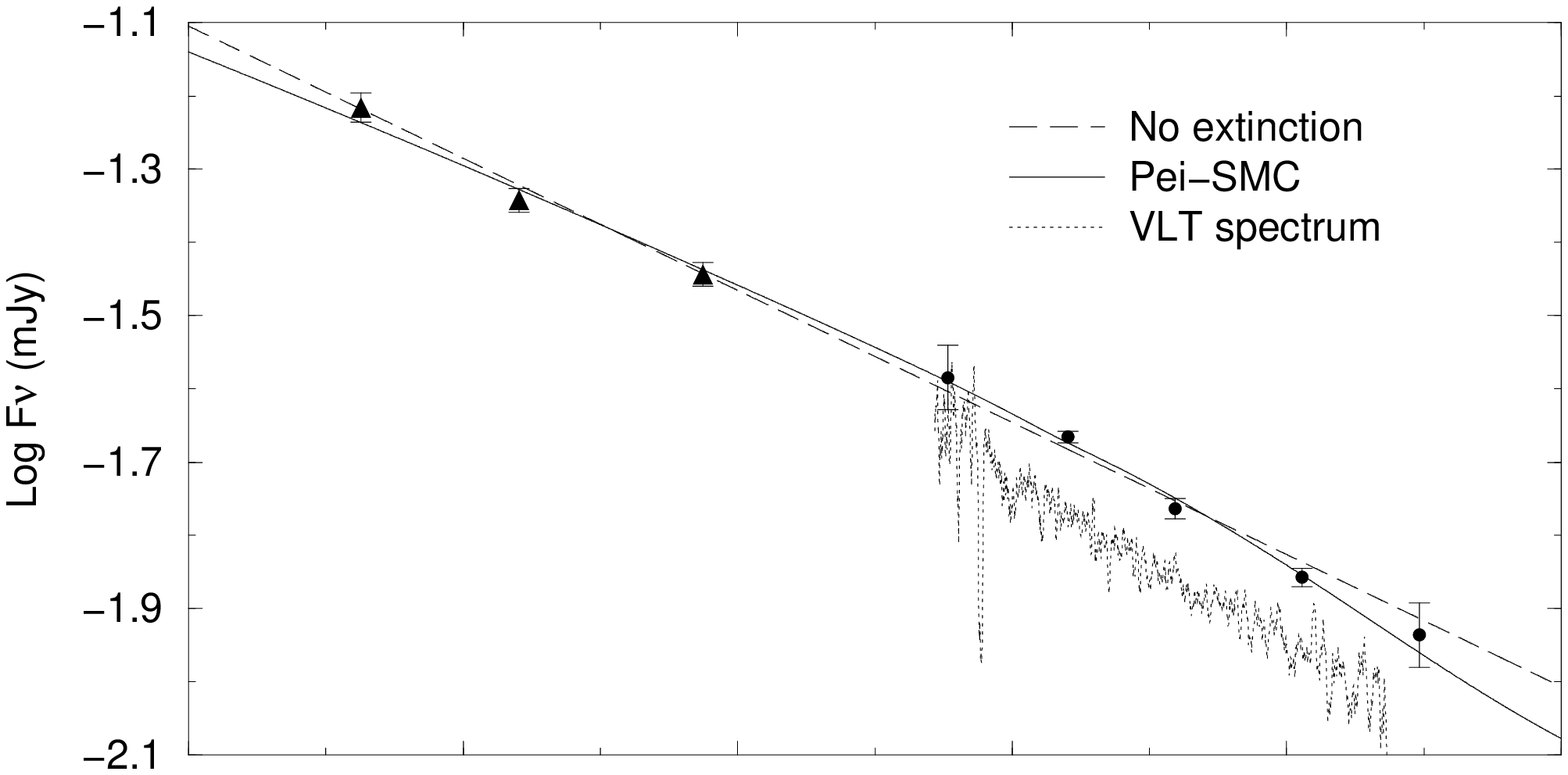}}
\resizebox{\hsize}{!}{\includegraphics[angle=270]{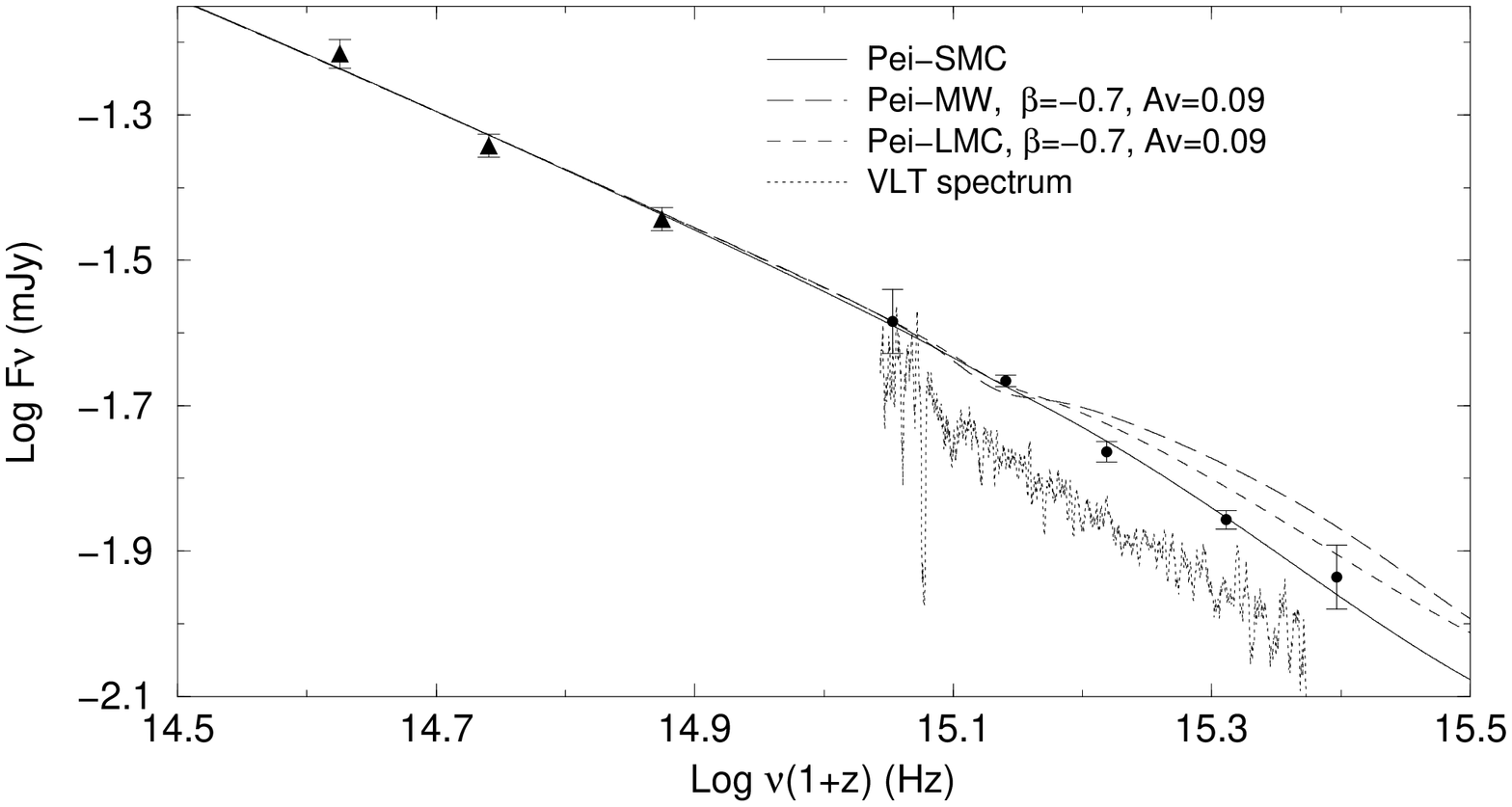}}
 \hfill
\caption{Spectral energy distribution of the OT on 2000 March
  4.39--4.55 UT ($\log \nu - \log \mathrm{F}_\nu$). The frequency scale has
  been  redshift  corrected to rest  frame  values assuming $z=2.0404$ (see
  Sect.~\ref{SECTION:redshift}).   The  data  points  include   our optical
  (UBVRI) and near  infrared (JHK) data  (see Table~\ref{TABLE:photometry})
  from March 4.5 UT.  In the plots, a range  of extinction models have been
  fitted to the data, cf.\  the text and Table~\ref{TABLE:SED}.  {\it Upper
    panel\/}:  UBVRIJHK photometry  with  the  VLT  spectrum (dotted line)
  of March   5.4
  included (the offset  from the photometry is due  to the fading of the OT
  between March 4.5  and 5.4).  The solid line represent the fitted SED when
  the emperical \object{SMC} extinction law is applied. A fit (longdashed)
  assuming no extinction of the host is shown for comparison.
  {\it Lower panel\/}:
  here is showed the effects on the \object{LMC} and
  \object{MW}-like SED fits when reasonable values  of $\beta$ and A$_{V}$
  are considered. The solid line, as in the upper panel, shows the SED
  when a \object{SMC}-like extinction law is fitted. Then, the values
  obtained of $\beta$ and A$_{V}$ ($\beta=0.7$, A$_{V}=0.09$) are applied
  to plot power-law afterglow SEDs with \object{MW} (long dashed) and
  \object{LMC}-like (dashed) extinction.
  {\it Circles\/}: USNO UBVRI-photometry,
  {\it Triangles}: UKIRT JHK-photometry.}
\label{FIGURE:SED}
\end{figure*}

The wide wavelength coverage obtained around 2000 March 4.5 UT allows us to
construct the multi-wavelength spectrum of the afterglow, by using our USNO
and UKIRT data. Most fireball models (e.g.\ Sari et~al.~\cite{SPN1998};
Piran~\cite{P1999}; M{\'e}sz{\'a}ros~\cite{M1999} and references therein)
and observations of previous afterglows suggest a power-law Spectral Energy
Distribution (SED). However, a global fit to the broadband SED of
\object{GRB~000301C} by
Berger et al.~(\protect\cite{BSF2000}) demonstrates that the mm--optical
range cannot be described by a single power-law. Thus, we have only
considered wavelengths shorter than IR when fitting the SEDs presented in this
section. This gives eight measurements in the period March 4.39--4.55 UT,
namely UBVRIJHK. The eight measurements, plotted against $\log \nu$
together with the VLT spectrum in Fig.~\ref{FIGURE:SED}, allow us to
constrain the SED of the OT at this epoch.

As extinction is   highly wavelength dependent,  a progressive deviation
from a pure power-law fit can be explained as  being due to dust extinction
in the GRB host galaxy.  In order to test this possibility,  we first fit a
power-law to  the data in the NIR--optical   range with the  addition of an
intrinsic extinction law,  i.e.,   using the expression  $F_{\nu}   \propto
\nu^{\beta}\times 10^{(-0.4 A_{\nu})}$,  where A{$_\nu$} is  the rest frame
extinction in magnitudes at the rest frame frequency $\nu$. Four extinction
laws have been applied in order to establish a relationship between A$_\nu$
and  $\nu$.  Following Rhoads \& Fruchter  (\cite{RF2001}),  we have fitted
the extinction curves for the Galaxy and Magellanic Clouds published by Pei
(\cite{P1992}).   Pei  (\cite{P1992})   provides extinction  laws  for  the
\object{Milky Way}  (\object{MW}),   the  \object{Large  Magellanic  Cloud}
(\object{LMC}),  and  the  \object{Small Magellanic   Cloud} (\object{SMC})
based on their different proportions  in the dust-to-gas ratio (1:1/5:1/10)
and in the  abundance of heavy  elements (1:1/3:1/8).  The most significant
difference is the  sequential  change in the strength   of the  2175  {\AA}
extinction feature, prominent     for  the \object{MW},   moderate in   the
\object{LMC}, and  nonexistent for the \object{SMC} extinction curve\@.  It
is important to   note that for the   redshift of \object{GRB~000301C}   this
extinction feature falls in the observed R-band. So, the presence of this
feature would result in a clear decrease in the R-band flux compared
to the I and
V-bands. The \object{MW}  extinction curve  requires   an equal amount   of
graphite and silicate  grains, while the  \object{SMC} extinction curve can
be explained by silicate grains only, with  the \object{LMC} extinction law
as an intermediate stage.  

We have applied  these four extinction  laws to our  data in order to infer
qualitative information about the dust-to-gas ratio, the abundance of heavy
elements and   the   composition  of  the    dust in the   host  galaxy  of
\object{GRB~000301C}. We leave A$_{\nu}$ as  a free parameter, so fitting a
function  like $F_{\nu}  \propto   \nu^{\beta}\times 10^{(-0.4   A_{\nu})}$
allows us to determine $\beta$ and A$_{\nu}$  simultaneously. The values of
$\chi^2$ are displayed in  Table~\ref{TABLE:SED}  for each case.

A pure   power-law fit ($F_{\nu}  \propto \nu^{\beta}$)  to  the eight data
points  (after   correction for  Galactic  extinction)  leads  to a reduced
$\chi^2$  of 1.69 (see  Table~\ref{TABLE:SED}), making an acceptable
description of the NIR--optical range of the SED.
However,   the fit can   be improved if a  modest amount of
extinction is introduced. This is because the  SED is slightly bending down
towards higher  frequencies  (see  Fig.~\ref{FIGURE:SED}).  The eight  data
points show that there is not any presence of a redshifted 2175 {\AA}
absorption bump in the R-band  at all. In short, the near-IR
SED of \object{GRB~000301C} can be described as a curved power-law but with
no broad absorption features.

As  expected  by  the  lack of the   absorption bump  in  the  R-band,  the
\object{MW} and the \object{LMC} extinction laws are completely inconsistent
with our data.  In fact, both fits imply an unphysical negative extinction
(see Table~\ref{TABLE:SED}). This is because the  R-band flux is slightly
over the linear interpolation between the I-band and  V-band fluxes (in  a
Log-Log space), and both of these two extinction laws fit
the 2175  {\AA} bump as an
emission feature instead of an absorption bump.  To illustrate the problem
with  the \object{MW} and \object{LMC} extinction curves, we have, in the
lower  panel  of  Fig.~\ref{FIGURE:SED}, plotted the   effect  of having the
\object{MW} and \object{LMC}  extinction laws  with the parameters  derived
for the \object{SMC}  extinction law  ($\beta$=-0.7, A$_V$=0.09). As  seen,
the shapes of both SEDs are incompatible with our UBVRIJHK measurements.

The quality  of   the   \object{MW},  \object{LMC} and   \object{SMC}   and
unextincted SEDs  can also be compared checking the flux predicted at
250~$\times$~(1+z)~GHz (rest-frame), where Berger et
al.~(\protect\cite{BSF2000}) reports a flux of 2.1 $\pm$ 0.3 mJy at 250 GHz
on March 4.29 UT.
Extrapolating the four extinction curves, we
obtain the following fluxes in increasing  order; 5.0 mJy (\object{SMC}),
18.1 mJy (No extinction), 33.9 mJy (\object{MW}) and 51.6 mJy
(\object{LMC}). As with the NIR--optical range, the \object{SMC} extinction
provides the most reasonable results. The actual measured flux in mm is
below the value predicted by the mm--optical extrapolation, because the
pure power-law  assumption is  not  correct in the mm-NIR spectral
range and an additional curvature effect  is present in  the SED, as
demonstrated by Berger
et al.~(\protect\cite{BSF2000}) (see their Fig. 1). Thus, the
value  of A$_V = 0.09 \pm 0.04$ for the SMC-fit given  in
Table~\ref{TABLE:SED} should be taken as a good indication of the
real extinction, although strictly
speaking it is just an upper limit.

In conclusion, the featureless SMC extinction law provides the best fit to
our data, improving the quality of the fit obtained for an unextincted
afterglow (see Table~\ref{TABLE:SED}). It is interesting to note the dramatic
dependence of the quality of fit on  the existence of the 2175 {\AA}
absorption bump. Extinction laws with high moderate dust-to-gas ratios
that produce such an absorption feature do not provide good fits to our
data points. Therefore, the spectral energy distribution of
\object{GRB~000301C}  supports a
scenario where the host is in the early stages of chemical enrichment.

\begin{table}[t]
\begin{center}
\caption{Result of fitting different extinction law models to
observations of the afterglow on 2000 March 4.5 UT\@. The extinction
curves include the models by Pei (\protect\cite{P1992}) for the
\object{Milky Way} (\object{MW}),
the \object{Large Magellanic Cloud} (\object{LMC}) and the
\object{Small Magellanic Cloud} (\object{SMC})  
(see Fig.~\ref{FIGURE:SED} and text for further details).}
\label{TABLE:SED}
\begin{tabular}{lccc}

\hline
                    &  $\chi^2$/DOF &  $\beta$              & A$_V$   \\
                    &               &                       & mag        \\
\hline
 No extinction     &     1.69     & $-0.90 \pm$ 0.02  &   0        \\
\hline
\object{MW}, Pei (1992)     &               &                    &  $<0$ \\
\object{LMC}, Pei (1992)    &               &                    &  $<0$ \\
\object{SMC}, Pei (1992)    &     0.91      & $-0.70 \pm$ 0.09  &  0.09 $\pm$ 0.04 \\
\hline
\end{tabular}
\end{center}
\end{table}

The power-law + extinction fits to the SED in the NIR--optical range 
allow us to predict the UV flux at the spectral range of MAMA-HST when the UV
spectrum was obtained at 2000 March 6.375 UT (Smette et al.~\cite{SFG2000})
assuming that the shape of the SED has not changed between the two
epochs. This assumption is supported by the imaging data (Sect. 6.3).
First, we consider the best fit to the SED at 2000 March 4.39--4.55 UT 
and calculate the flux at March 4.39-4.55 UT at 3000~{\AA}. Then,
making use of the light curve models (presented in Table~\ref{TABLE:fits}),
we estimate the value of the flux at 3000~{\AA} for March 6.375 UT.
The predicted flux ranges from 5.9 $\times 10^{-18}\,$
erg$\,$cm$^{-2}\,$s$^{-1}$~{\AA}$^{-1}$ to 7.9 $\times 10^{-18}\,$
erg$\,$cm$^{-2}\,$s$^{-1}$~{\AA}$^{-1}$,
depending on the light curve model.
A final analysis (Smette et al.~\cite{SFG2001}) of the MAMA-HST
data revealed a flux of $\sim7.3^{+0.8}_{-1.8}~10^{-18}$ erg
cm$^{-1}$ consistent with our extrapolation.

\subsection{Evolution of the spectral energy distribution}
\label{SECTION:SEDres}
\begin{figure}
\resizebox{\hsize}{!}{\includegraphics{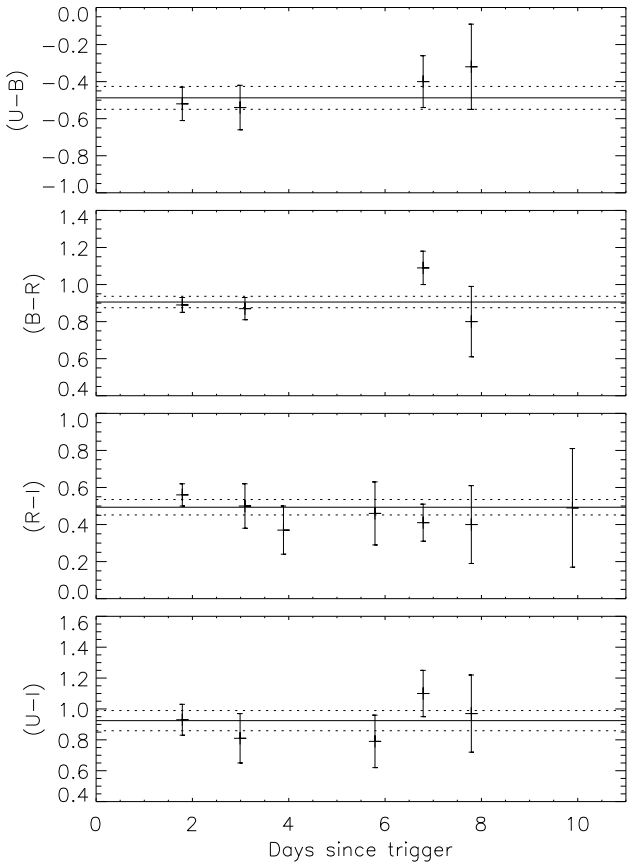}}
\caption{Fitting the colours from the optical photometry from 2000 March 3 to March 11 UT, assuming achromatic evolution. The solid line indicates the best fit, with the dashed lines representing the error-range. The photometry is from Table~\ref{TABLE:photometry}. The results of the fitting is presented in Table~\ref{TABLE:colours}.}
\label{FIGURE:colours}
\end{figure}

 From our UBRI photometric data (presented in Table~\ref{TABLE:photometry}) we have multi-band optical coverage from 2 to 10 days after the burst-trigger (on March 1.41 UT). When analysing the colours, we find that the simplest reliable fit is for constant colours. Thus we find no evidence for optical chromatic evolution for the afterglow during the period of observations (see Fig.~\ref{FIGURE:colours}). For these constant fits, we obtain the values presented in Table~\ref{TABLE:colours}. These values are not corrected for Galactic or intrinsic extinction. \\

\begin{table}
\begin{center}
\caption{Best fits for the colours of the OT from 2000 March 3 to 11 UT, assuming an achromatic evolution. V$-$R is for March 4.4 only. P($\chi^2$) is the probability to obtain a lower value of $\chi^2$/DOF for the given model (constant colour). Colours are not corrected for galactic or intrinsic extinction. See also Fig.~\ref{FIGURE:colours}.}
\label{TABLE:colours}
\begin{tabular}{lrccc}
Colour & Value &    $\sigma$ &    $\chi^2$/DOF &    P($\chi^2$) \\
\hline
U$-$B &   $-0.49$&    0.06 &     0.41 &    0.26 \\
B$-$R &     0.91 &    0.03 &     1.67 &    0.83 \\
R$-$I &     0.49 &    0.04 &     0.51 &    0.20 \\
B$-$I &     1.43 &    0.05 &     0.33 &    0.14 \\
U$-$I &     0.92 &    0.07 &     0.64 &    0.36 \\
U$-$R &     0.40 &    0.05 &     1.35 &    0.75 \\
V$-$R &     0.44 &    0.05 &        - &       - \\
\hline

\end{tabular}
\end{center}
\end{table}

\subsection{The light curve\label{SECTION:light_curve}}

        According to the simple fireball model the optical afterglow
should follow a power-law decay, $F_{\nu} \propto \nu^{\beta}
t^{\alpha}$ (Sari et~al.~\cite{SPN1998}). However, a single power-law
is excluded at more than the $99.9$\% confidence level.  The
parameters for this power-law are given in Table~\ref{TABLE:fits}.
The photometry suggests that the optical afterglow follows a shallow
power-law decay for the first few days and then steepens.  This
behaviour has been seen previously in \object{GRB~980519} (Jaunsen
et~al.~\cite{J2001}), \object{GRB~990123} (Kulkarni
et~al.~\cite{KDO1999}), \object{GRB~990510} (Harrison
et~al.~\cite{HBF1999}), \object{GRB~991208}
(Castro-Tirado et~al.~\cite{CT2001}) and \object{GRB~000926}
(Fynbo et~al.~\cite{FGD2001} and is
predicted by many models for gamma-ray bursts (see below).  Sagar
et~al.~(\cite{SMP2000}) report that there are seven components to the
R-band light curve. Here we are primarily interested in the
overall structure of the light curve, not the structure at small time
scales.  Therefore, we fit a broken power-law of the form,
\begin{equation}
\label{EQUATION:broken_power_law}
f_{\nu}(t) = \left \{
        \begin{array}{lll}
                f_{\nu}(t_b) {\left(\frac{t}{t_b}\right)}^{\alpha_1}, &
                \mathrm{if}  &  t \le t_b \\
                f_{\nu}(t_b) {\left(\frac{t}{t_b}\right)}^{\alpha_2}, &
                \mathrm{if}  &  t \ge t_b,
        \end{array}
             \right.
\end{equation}
to the UBRI data presented in Table~\ref{TABLE:photometry}.  The flux,
in $\mu$Jy, at time $t$ days after the burst is denoted by
$f_{\nu}(t)$.  The time of the break in the decay is denoted $t_b$.
The slope before the break is $\alpha_1$, and the slope after the
break is $\alpha_2$.  The flux at the time of the break is
$f_{\nu}(t_b)$.  We used CERN's {\sc Minuit} function minimization
package, and a chi-square minimization scheme, to simultaneously solve
for the four free parameters ($\alpha_1$, $\alpha_2$, $t_b$, and
$f_{\nu}(t_b)$) and their formal 1-$\sigma$ errors in the fit for each
parameter.

        The data was corrected for Galactic reddening and extinction
before the fits were made.  No corrections were made for reddening or
extinction in the host galaxy.  The photometry was transformed to the
R band using the colours given in Table~\ref{TABLE:colours} and
then converted to units of flux using a photometric zero point of
$f_{\nu,0} = 3.02 \times 10^{-20}$ erg cm$^{-1}$ s$^{-1}$ Hz$^{-1}$
(Fukugita et~al.~\cite{FSI1995}).

\begin{figure}
\resizebox{\hsize}{!}{\includegraphics{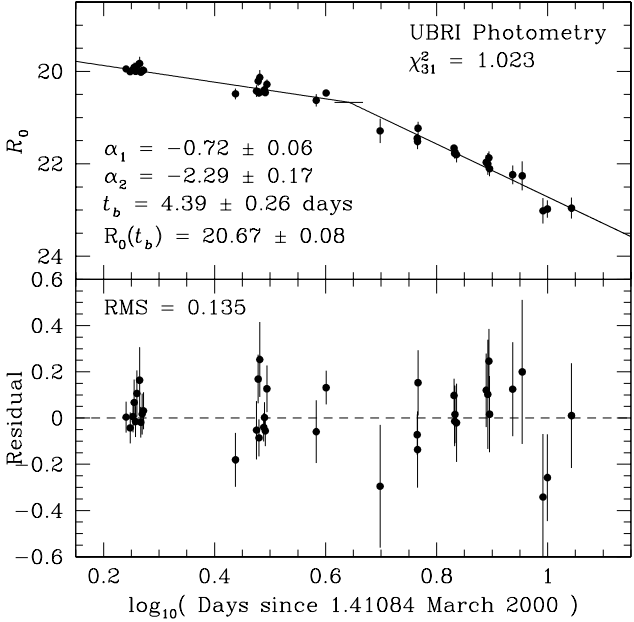}}
\caption{The upper panel shows the UBRI photometry from
Table~\ref{TABLE:photometry} and the best-fitting broken power-law fit
to the data.  The photometry was offset to the R band using the
colours given in Sect.~\ref{SECTION:SEDres} (see the text).  The
horizontal line at the location of the break shows the 1-$\sigma$
uncertainty in the time of the break.  The lower panel shows the
residuals of the fit in the sense $\mathrm{R}_\mathrm{obs} -
\mathrm{R}_\mathrm{fit}$.  The uncertainties in the residuals are the
uncertainties in the photometry.}
\label{FIGURE:ubri_decay}
\end{figure}

        The best fit to the combined UBRI photometry is listed in
Table~\ref{TABLE:fits}, and shown in Fig.~\ref{FIGURE:ubri_decay}.
 To test the sensitivity of the results to the fitting function, we also
fit our data with the smooth function used by Stanek et~al.~(\cite{SGK1999},
 their Eq.~1) on \object{GRB~990510}. The results are
given in Table~\ref{TABLE:fits}. The broken power-law gives the
smallest chi-square value, and the errors in the individual parameters
are smaller for the broken power-law fit than they are for the smooth function.
The correlation coefficient between $t_b$ and $\alpha_1$ is
$-0.39$ and the coefficient between $t_b$ and $\alpha_2$ is $-0.84$.
The broken power-law fit is consistent with the data at the 43\%
confidence level.  
Even though we, from the theory of fireballs, would expect that the light-curve
evolution is a smooth function, we find in the case
of \object{GRB~000301C}, in agreement with Berger et al.~(\cite{BSF2000}),
that the broken power-law provides the most reliable fit.
 Additionally, a broken power-law provides the most reliable metod
of determining the time when the decay of the light has steepened,
and thus is a useful way of parameterising the data.

\begin{figure}
\resizebox{\hsize}{!}{\includegraphics{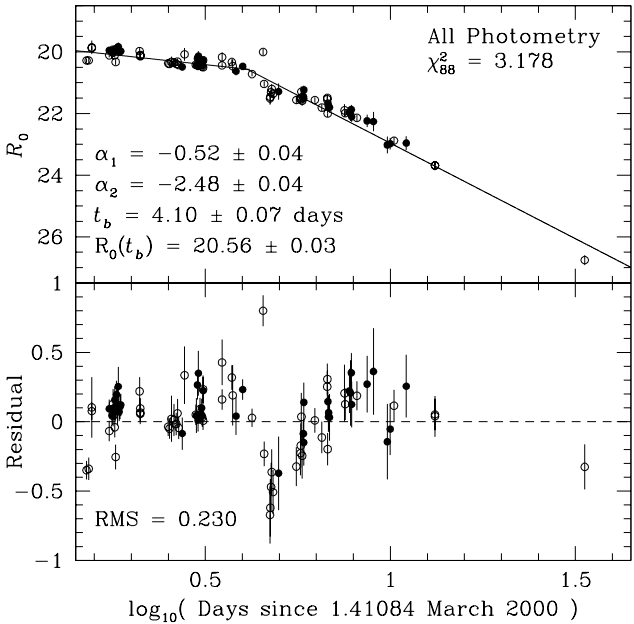}}
\caption{This Figure is analogous to Fig.~\ref{FIGURE:ubri_decay}, but instead shows the data from Table~\ref{TABLE:photometry} (filled circles) together with all the UBRI photometry presented in the literature (Rhoads \& Fruchter~\cite{RF2001} and references therein) (open circles)
and with the best-fitting broken power-law fit to this data.}
\label{FIGURE:all_decay}
\end{figure}

        We combined all our UBRI-data (Table~\ref{TABLE:photometry}) with those from the literature. Fig.~\ref{FIGURE:all_decay} shows the best-fitting broken
power-law for all of the UBRI data in the literature
(Sagar et al.~\protect\cite{SMP2000} and references therein)
and Table~\ref{TABLE:photometry}.  This data was
shifted to the R band in the manner described above.  The parameters
of the fit are shown in Fig.~\ref{FIGURE:all_decay} and are not
significantly different from the parameters of the broken power-law
that was fit to our data (see Table~\ref{TABLE:fits}).
The conspicuous short-term behaviour of the light-curve has been detailed
by Masetti et al.~(\cite{MBB2000}), Sagar et al.~(\cite{SMP2000})
and Berger et al.~(\cite{BSF2000}). Garnavich, Loeb \& Stanek
(\cite{GLS2000}) find that the variation of the lightcurve can be
interpreted as a microlensing event, peaking about 3.5 days after the burst,
superposed on a power-law broken at $t_b = 7.6$ days. 
This superposed event peaks at a more sparsely sampled period in our data,
coinciding partly with where we identify the break. Thus it is not
possible, from our data, to further constrain the existence of such an event.
We choose here to work only with the data presented in
Table~\ref{TABLE:photometry} as it represents a consistently derived set.



\begin{table}
\begin{center}
\caption{The parameters of the best-fitting functions to the optical
decay of \object{GRB~000301C}.  The number of degrees-of-freedom (DOF)
in each fit is the number of data points minus the number of
parameters.  The number of parameters is the number of free parameters
in each model plus the number of colours that the data was adjusted
with, in order to bring it into the R band.}
\smallskip
\label{TABLE:fits}
{\scriptsize
\begin{tabular}{cccc}
\hline
\hline
               & Single Power Law  &  Broken Power Law & Smooth Function \\
\hline
\vspace{0.1em}
$\alpha$       & $-1.22 \pm 0.07$  &     ~$\cdots$~    &   ~$\cdots$~ \\
$\alpha_1$     &   ~$\cdots$~      & $-0.72 \pm 0.06$  &  $-0.55 \pm 0.14$ \\
$\alpha_2$     &   ~$\cdots$~      & $-2.29 \pm 0.17$  &  $-3.53 \pm 0.47$ \\
$t_b$          &       0           &  $4.39 \pm 0.26$  &   $5.99 \pm 0.79$ \\
$f_{\nu}(t_b)$ & $70.46 \pm 0.56$  & $16.23 \pm 1.13$  &   $8.24 \pm 2.05$ \\
$R_0(t_b)$     & $19.08 \pm 0.01$  & $20.67 \pm 0.08$  &  $21.41 \pm 0.27$ \\
$\chi^2/$DOF   &       4.499       &       1.023       &        1.090 \\
DOF            &      33           &      31           &       31 \\
\hline
\hline
\end{tabular}
}
\end{center}
\end{table}

\section{Discussion}
\label{SECTION:discussion}

\subsection{Interpretation of the light curve}
\label{SECTION:interpretation}

The fit to the multi-colour light curves shows a break at
$t_b=4.39\pm0.26$ days, with the light curve steepening from $\alpha_1
= -0.72\pm0.06$ to $\alpha_2 = -2.29\pm0.17$, i.e., by $\Delta \alpha
= \alpha_1-\alpha_2=1.57\pm0.18$.  A broken light curve can arise in a
number of circumstances: {\it i)\/} If the frequency separating fast
cooling electrons from slow cooling ones moves through the optical at
$t_b$, the resulting light curve would steepen by $\Delta \alpha \sim
0.25$ (Sari et~al.~\cite{SPN1998}).  {\it ii)\/} The light curve may
also steepen if a spherical fireball slows down to a non-relativistic
expansion (Dai \& Lu \cite{DL1999}), resulting in $\Delta
\alpha=-(\alpha_1+3/5)=~0.12$ for our value of $\alpha_1$.
{\it iii)\/} If the outflow is
collimated with a fixed opening angle, the break in the light curve
occurs when the relativistic beaming of the synchrotron radiation
becomes wider than the jet opening angle (M\'esz\'aros \& Rees
\cite{MR1999}). In this case the break is a geometrical effect and the
steepening is $\Delta \alpha =3/4$.  {\it iv)\/} If the afterglow
arises in a sideways expanding jet, the steepening will be $\Delta
\alpha =(1-\alpha_1/3)=1.24$ (Rhoads~\cite{R1999}) for our value of $\alpha_1$.
The above estimates all assume a constant mean density distribution
of the ambient medium. We note that collimated outflows in general
result in faster decaying light curves than the spherically
symmetric ones. If the mean density distribution is not constant,
e.g. it has a stellar wind density profile, the light curves also
decay faster, but the break will be less pronounced
(Panaitescu et~al.~\cite{PMR1998}).

Based on the light-curve properties alone, the model that best fits the 
observations is that of a sideways expanding jet in an ambient medium with a 
constant mean density distribution. In that interpretation, the observed light 
curve indices imply an electron energy distribution index of $p=2.13\pm0.09$ 
that results in a theoretical spectral index of 
$\beta=-(p-1)/2=-0.56\pm$0.05.
This is in agreement with the spectral 
index of $\beta=-0.70\pm$0.09 inferred from our spectroscopic
observations when correcting for
extinction in the host galaxy (Sect.~\ref{SECTION:SpectroscopyRes}),
independently strengthening the described model for the afterglow. 

With a combined fluence in the 25 -- 1000 KeV range of
$4\times 10^{-6}$ erg cm$^{-2}$, and a redshift of
$z=2.0404$, the isotropic energy release of GRB~000301C is $E = 4.6
\times 10^{52}$ erg.  Following Rhoads (\cite{R1999}),
the energy estimate and the light curve break
time, $t_b=4.39 \pm 0.26$ days, implies a jet opening angle, at that
time, of $\theta \approx 15^\circ n^{1/8}$, where $n$ is the number
density of the ambient medium (in units of cm$^{-3}$), and the break is assumed to occur when the opening angle equals the bulk Lorentz factor, $\theta = \Gamma$.

        This interpretation is similar to that of GRB~990510 which was
almost 5 times more energetic, but had a jet opening angle of
approximately $5^\circ$, leading to an earlier break in the light
curve. The best fit to GRB~990510 was a smooth function (e.g.\ Stanek
et~al.~\cite{SGK1999}; Harrison et~al.~\cite{HBF1999}; Holland et
al.~\cite{HBH2000}), as compared to a broken power-law in the case of
GRB~000301C.

\subsection{The host galaxy}
\label{SECTION:host}

   The host galaxy appears to be very faint. There is no evidence
for any extended emission from a host galaxy in any of the data
presented in this paper. Deep images obtained with
HST+STIS about 1 month after the GRB indicate that any host galaxy
must be R$\geq$27.8$\pm$0.25 (Fruchter et~al.~\cite{FSG2000}).
Hence, we can safely conclude that the host galaxy of \object{GRB~000301C} is
very faint compared to other known populations of galaxies at high redshifts. 
From fitting extinction curves to our photometry, we have found
evidence for some extinction ($\mathrm{A}_V$ $\sim$ 0.1) in the 
host galaxy. The A$_{V}$ derived from the best fit corresponds to an 
absorption at rest-frame 1500~{\AA} of about 0.4 magnitudes. 

For comparison, Lyman Break Galaxies (LBGs) at slightly higher redshifts
(z $\sim$ 2.5--3.5) on average have values of extinction at
 rest frame wavelength
1500 {\AA} of approximately 1.7 magnitudes, and, in rare cases up to 5
magnitudes (Steidel et~al.~\cite{SADP1999}). Hence, the faint optical
appearance of the host galaxy relative to the star-forming LBGs at
high redshift is most likely not imposed by massive extinction, but is
rather due to a lower overall star formation rate of the host galaxy.

As the host galaxy furthermore has a very high \ion{H}{i} column 
density, log(N\ion{H}{i})=21.2$\pm$0.5 as derived from the Ly$\alpha$
absorption feature and supported by the strong Lyman break 
(Smette et al.~\cite{SFG2000}), it is interesting to compare
with the population of galaxies identified as 
Damped Ly$\alpha$ Absorbers (Wolfe et al.~\cite{WTS1986}) in the
spectra of background QSOs. These galaxies have \ion{H}{i} 
column densities higher than log(N\ion{H}{i})=20.3. Based on 
the luminosity function of LBGs and the typical impact parameters of 
DLAs, Fynbo et al.~(\cite{FMW1999}) show that the majority
of DLAs at $z=3$ must be 
fainter than the current flux limit for LBGs of R=25.5 and that there 
hence is a very abundant population of galaxies fainter than the LBG 
flux limit. A similar conclusion has been reached by Haehnelt et al.
(\cite{HSR2000}). 
The dust-to-gas ratio towards the line-of-sight of {\object GRB~000301C} gives
a value of A$_V$/N($\ion{H}{i}) \leq {0.1}/1.6\times 10^{21}$ cm$^{2}
= 0.6\times 10^{-22}$ cm$^{-2}$. This upper limit is, within errors,
consistent with the expected A$_V$/N($\ion{H}{i})$
for DLAs. The corresponding value for the Milky Way is $2\times
10^{-22}$ cm$^{-2}$ (Allen~\cite{ALL2000}).

It is still uncertain what fraction of the integrated star 
formation rate at high redshift is accounted for by the LBGs and 
what fraction has to be accounted for by galaxies further down the 
luminosity function.  
The relative occurrence of GRBs in a given population of galaxies is 
expected to be proportional to its relative contribution to the total star 
formation rate (Totani et al.~\cite{T1997};
Wijers et al.~\cite{WBB1998};
Mao \& Mo~\cite{MM1999}; Blain \& Natarajan \cite{BN2000}).
However, so far only one GRB 
(\object{GRB~971214} at $z=3.418$) is confirmed to have occurred in
a galaxy similar to the faint members of the LBGs selected in ground
based surveys (Odewahn et al.~\cite{ODK1998}).
 The fact that \object{GRB~000301C}  
occurred in an intrinsically very faint galaxy and that most GRBs 
with identified OTs have occurred in L$^*$ or sub-L$^*$ galaxies, 
suggest that a large fraction of total star formation at high redshift occurs
in a population of galaxies that is further down the luminosity function 
than the bright LBGs found in ground based surveys and that is likely to have
a large overlap with the DLAs.

\section{Conclusion}
\label{SECTION:conclusion}

\object{GRB~000301C} is so far the GRB of shortest duration, for which a 
counterpart has been detected.
The high-energy properties of the burst are consistent
with membership of the short-duration class
of GRBs, though \object{GRB~000301C} could belong 
to the proposed intermediate class of GRBs or the extreme short
end of the distribution of long-duration GRBs.
Our VLT-spectra show that \object{GRB~000301C} occurred at a redshift
of 2.0404$\pm$0.0008. The light curve of the optical transient is
well-fitted by a broken power-law and it is consistent with being
achromatic. From the light-curve properties we find that the best model
for \object{GRB~000301C} is that of a sideways expanding jet in an ambient
medium of constant density.
This interpretation is further supported by the achromatic 
light-curve evolution, and by the agreement between the theoretically 
predicted and observationally derived spectral indices. The spectral energy 
distribution at March 4.5
reveals \object{SMC}-like extinction in the host galaxy at a level of 
$\mathrm{A}_V < 0.10$, which is significantly lower than for
the strongly star-forming LBGs. Hence, the extreme faintness of the
host galaxy indicates a low overall star-formation rate in the 
host galaxy, raising the possibility that the host may be a chemically 
less evolved, relatively low-luminosity galaxy containing SMC-type 
dust. We argue that there may be a connection between the host galaxy of 
\object{GRB~000301C} and DLAs, suggesting 
that substantial star-forming activity at high redshift takes place in 
relatively faint galaxies. Future studies of high redshift GRBs will further 
help explore this connection.

\section*{Acknowledgments}
We wish to thank our anonymous referee for providing many helpful comments and suggestions.
Support for this programme by the director of the Nordic Optical
Telescope, professor Piirola, is much appreciated.
We also acknowledge the assistance given by the ESO service
observing team. J.\ Gorosabel acknowledges the receipt of a Marie Curie
Research Grant from the European Commission. J.\ Hjorth acknowledges
support from the Danish Natural Science Research Council (SNF).
B.\ Thomsen acknowledges support from the Danish Natural Science
Research Council for funding the Danish Centre for Astrophysics with
the HST. G.\ Bj\"{o}rnsson acknowledges support from the Icelandic
Research Council and the University of Iceland Research Fund.
I.\ Burud is supported by P{\^o}le d'Attraction Interuniversitaire,
P4/05 \protect{(SSTC, Belgium)}.
K.\ Hurley acknowledges support for Ulysses operations under JPL Contract
958056, for IPN operations under NASA LTSA grant NAG5-3500, and for
NEAR operations under the NEAR Participating Scientist program.  We are
grateful to R.\ Gold and R.\ McNutt for their assistance with the NEAR
spacecraft.  R.\ Starr is supported by NASA grant NCC5-380.  We are indebted
to T.\ Sheets for her excellent work on NEAR data reduction.  Special
thanks also go to the NEAR project office for its support of
post-launch changes to XGRS software that made these measurements
possible.  In particular, we are grateful to John R.\ Hayes and Susan E.\
Schneider for writing the GRB software for the XGRS instrument and to
Stanley B.\ Cooper and David S.\ Tillman for making it possible to get
accurate universal time for the NEAR GRB detections.
The data presented here have been taken using ALFOSC, which is owned by the
Instituto de Astrofisica de Andalucia (IAA) and operated at the Nordic
Optical Telescope under agreement between IAA and the NBIfAFG of the
Astronomical Observatory of Copenhagen.
Additionally, the availability of the GRB Coordinates Network (GCN) and
BACODINE services, maintained by Scott Barthelmy, is greatly acknowledged.
We acknowledge the availability of POSS-II exposures, used in this work;
The Second Palomar Observatory Sky Survey (POSS-II) was made by the California
Institute of Technology with funds from the National Science Foundation,
the National Aeronautics and Space Administration, the National
Geographic Society, the Sloan Foundation, the Samuel Oschin Foundation,
and the Eastman Kodak Corporation.

\end{document}